# Implementing non-Abelian Hatano-Nelson model in electric circuits

Xiangru Chen[1†], Jien Wu[2†], Xingyu Chen[1], Zhenhang Pu[1], Yejian Hu[1*], Jiuyang Lu[1], Manzhu Ke[1], Weiyin Deng[1*], and Zhengyou Liu[1,3*]

1Key Laboratory of Artificial Micro- and Nanostructures of Ministry of Education and School of Physics and Technology, Wuhan University, Wuhan 430072, China
2School of Physics and Electronic Science, Changsha University of Science and Technology, Changsha 410114, China
3Institute for Advanced Studies, Wuhan University, Wuhan 430072, China

†These authors contributed equally to this work.
*Corresponding author. Email: huyejian@whu.edu.cn; dengwy@whu.edu.cn; zyliu@whu.edu.cn

Non-Hermitian systems generally host complex spectra that bring unique spectral topologies, leading to the spectral braiding and non-Hermitian skin effect. The experimental exploration of non-Hermitian physics is mainly concentrated in artificial systems due to the flexibility in the introduction of the non-Hermiticity, but to date has focused only on the systems without gauge fields or with Abelian gauge fields. Here, we propose a non-Abelian Hatano-Nelson model with a nonreciprocal $U(2)$ gauge field. The gauge field induces two non-Hermitian phenomena: the first is the Hopf-link-shaped complex energy braiding, and the second is the bipolar skin effect arising under the non-Abelian condition. The non-Abelian Hatano-Nelson model is implemented in electric circuits, and the Hopf-link-shaped admittance spectra and bipolar skin admittance modes are observed. Our work enriches the experimental non-Hermitian physics, and provides an approach to designing multifunctional non-Hermitian devices.

In topological materials, the non-trivial evolution of eigenvectors gives rise to a rich tapestry of topological phenomena [1]. Once non-Hermitian perturbations are introduced, a natural consequence is that the eigenvalue spectra expand from real values to complex ones [2]. The complex spectra play a unique and pivotal role in the emergence of topological phenomena, adding another dimension to our understanding of these complex behaviors beyond the framework of eigenvector topology [3]. From the perspective of spectral topology, there are two notable and distinct phenomena. The first is the complex energy braiding characterized by braiding degree [4-6]. Rich knot and braiding structures of non-Hermitian bands have been proposed and observed in acoustic cavity [4], cavity optomechanical system [5], and photonic crystal [6]. The second is the non-Hermitian skin effect, which is characterized by spectral winding number [7-11]. At present, different types of skin effects have been achieved, such as the bipolar skin effect in one-dimension (1D) with long-range couplings [12-14], higher-order skin effect [15-17] and geometry-dependent skin effects [18-22] in higher-dimensions. To date, most studies on non-Hermitian physics have been limited to systems without gauge fields or with Abelian gauge fields [14,23-29].

In recent years, non-Abelian physics has attracted extensive attention, exhibiting numerous intriguing phenomena including non-Abelian anyons [30,31], non-Abelian topological charges [32-35], and non-Abelian mode braiding [36-39]. Particularly, non-Abelian gauge fields with spin or pseudospin degree of freedom play a central role in enabling the non-Abelian physics, and have been implemented in diverse physical systems [40-43]. It has been discovered that there are many topological phenomena associated with the non-Abelian properties of gauge fields, such as the re-emergence of the Hofstadter butterfly [44,45], the localized behavior of states [46,47], and the existence of Dirac points at special momentum points [48]. Given these advancements, exploring the interplay between non-Abelian gauge fields and non-Hermiticity is of great worth. Current theoretical studies on 1D non-Hermitian lattices with imbalanced non-Abelian couplings show that $SU(2)$ gauge fields generate rich non-Hermitian topologies, despite the absence of gauge fluxes in 1D [49,50]. However, such exotic non-Hermitian topologies induced by non-Abelian gauge fields remain experimentally unexplored.

In this work, we propose and realize a non-Abelian Hatano-Nelson model with a nonreciprocal $U(2)$ gauge field. Its complex spectra can not only braid as a Hopf link with braiding degree $|\nu| = 2$ but also guarantee the bipolar skin effect with opposite spectral windings arising under the non-Abelian condition. In traditional 1D models, long-range couplings are typically necessary to attain higher complex energy braiding with braiding degree $|\nu| > 1$ [6,51] or to realize the bipolar skin effect [12-14]. However, we can obtain these phenomena simultaneously by only introducing the nearest-neighbor non-Abelian couplings. Benefiting from mature electric technology, electric circuits are an excellent platform for realizing many exotic topological phenomena [52], including the Hermitian [53-57], non-Hermitian [58-62], and non-Abelian [41,42] topological states. We implement the lattice model in electric circuits, where the nonreciprocal non-Abelian matrix-valued couplings are achieved by different link configurations of electric elements. The complex spectrum braiding and the skin effect are experimentally confirmed by measuring spectra and eigenstates of the admittance matrix, respectively.

We construct a non-Abelian Hatano-Nelson model with a nonreciprocal $U(2)$ gauge field, as shown in Fig. 1(a). This model features on-site potential $t_0 \boldsymbol{d}_R \cdot \boldsymbol{\sigma}$ along with nonreciprocal matrix-valued couplings $t_L \boldsymbol{d}_L \cdot \boldsymbol{\sigma}$ and $t_R \boldsymbol{d}_R \cdot \boldsymbol{\sigma}$, where $\boldsymbol{d}_L$, $\boldsymbol{d}_R$ are real vectors of unit length with $\boldsymbol{d}_{L(R)} = \left(d_{L(R),x}, d_{L(R),y}, d_{L(R),z}\right)$, and $\boldsymbol{\sigma} = \left(\sigma_x, \sigma_y, \sigma_z\right)$. $t_L$ and $t_R$ represent the hopping amplitudes, and $t_0$ is the potential amplitude. The non-Hermitian Hamiltonian in real space is

$$\hat{H} = \sum_{n=1}^{N} a_n^\dagger t_0 \boldsymbol{d}_R \cdot \boldsymbol{\sigma} a_n + \sum_{n=1}^{N-1} \left(a_n^\dagger t_L \boldsymbol{d}_L \cdot \boldsymbol{\sigma} a_{n+1} + a_{n+1}^\dagger t_R \boldsymbol{d}_R \cdot \boldsymbol{\sigma} a_n\right), \qquad (1)$$

where $a_n^\dagger = \left[a_{n,\uparrow}^\dagger, a_{n,\downarrow}^\dagger\right]$ ( $a_n = \left[a_{n,\uparrow}, a_{n,\downarrow}\right]^T$ ) is the two-component creation (annihilation) operators with two pseudospins. $N$ is the total number of sites, $k$ is the dimensionless wave vector and the distance of nearest-neighbor sites is set to be a unit. The non-Abelian condition of gauge field is ensured by $[\boldsymbol{d}_L \cdot \boldsymbol{\sigma}, \boldsymbol{d}_R \cdot \boldsymbol{\sigma}] \neq 0$ of the nonreciprocal couplings, and the Abelian condition is $[\boldsymbol{d}_L \cdot \boldsymbol{\sigma}, \boldsymbol{d}_R \cdot \boldsymbol{\sigma}] = 0$. The corresponding Bloch Hamiltonian with periodic boundary condition (PBC) is

$$H = t_0 \boldsymbol{d}_R \cdot \boldsymbol{\sigma} + t_L e^{ik} \boldsymbol{d}_L \cdot \boldsymbol{\sigma} + t_R e^{-ik} \boldsymbol{d}_R \cdot \boldsymbol{\sigma}, \tag{2}$$

leading to the complex eigenvalues

$$E_\pm = \pm\sqrt{(t_0 + t_R e^{-ik})^2 + t_L^2 e^{2ik} + 2(t_0 + t_R e^{-ik}) t_L e^{ik} \boldsymbol{d}_L \cdot \boldsymbol{d}_R}$$

In this system, both the amplitudes $t_L$ and $t_R$ and the $U(2)$ matrices $\boldsymbol{d}_L \cdot \boldsymbol{\sigma}$ and $\boldsymbol{d}_R \cdot \boldsymbol{\sigma}$ of the matrix-valued couplings contribute to non-Hermiticity, resulting in complex spectra. In this work, we choose $\boldsymbol{d}_L = (0,0,1)$ and $\boldsymbol{d}_R = (1,0,0)$ as an example. These complex spectra can be regulated by the parameters $t_L$ and $t_R$, providing the possibility for achieving spectral topologies.

One spectral topology is the higher complex energy braiding with $|v| > 1$, where two energy bands can braid with each other and form a Hopf link. The spectral braiding property is a characteristic of a non-Hermitian system, and the topology of this braiding can be described by the braiding degree [6,51] defined as

$$v = \int_0^{2\pi} \frac{dk}{2\pi i} \frac{d}{dk} \ln \det\left(H - \frac{1}{2}\mathrm{Tr}\, H\right)$$

Figure 1(b) exhibits the phase diagram of braiding degree in the $t_L$-$t_R$ plane. There are three distinct phases with $v = 0$ and $v = \pm 2$. The phase boundaries described as $1 + (t_L^2 + t_R^2)[2\, t_R^2/(t_L^2 - t_R^2)^2 - 1] + 2\, t_R^2/(t_L^2 - t_R^2) = 0$, marking the emergence of exceptional points (Supplementary Material [63]). This braiding degree quantifies the number of times that two bands braid in the $E$-$k$ space, where the sign represents the handedness. For example, we chose a parameter point $P_1$ within area with $v = -2$, and plot its bands in the $E$-$k$ space, as shown in Fig. 1(c). Dashed lines on the $k = 0$ plane represent the projection of two energy bands, and arrows illustrate the winding direction as $k$ increases. One can see that energy bands form two nested loops, namely, a Hopf link, where the winding direction is clockwise. Braiding degree $v = 2$ can also produce a Hopf link but the winding direction is counterclockwise, as shown in Fig. 1(d). Braiding degree $v = 0$ indicates that a Hopf link will disappear, and the two bands form unlink and are unbraided. The formation of Hopf links indicates that the presence of non-Abelian gauge field can lead to more intricate braiding structures in 1D systems. Evolution of the complex energy braiding is provided in the Supplementary Material [63].

Another spectral topology is the bipolar skin effect under open boundary condition (OBC). The topology of skin effect, originating from the point gap of complex spectra, is described by the spectral winding number defined as $w = (-i/2\pi) \int_0^{2\pi} \ln \det[H - E_0] dk$ with $E_0$ being the reference energy. $w = -1$ ($w = 1$) has the clockwise (counterclockwise) spectral loop on the complex plane and guarantees the presence of right-localized (left-localized) skin effect for OBC eigenstates. In order to directly elaborate the localization of OBC eigenstates, we define a comparison function $\gamma$ as $\gamma = (D_L - D_R)/(D_L + D_R)$ with $D_{L,R} = \sum_{x \in L,R} |\psi(x)|^2$, where $L$ and $R$ represent the left and right boundary lengths, respectively. Specifically, here the OBC chain has 100 sites, and $L$ ($R$) contains sites 1-20 (81-100). Figure 1(e) depicts $\gamma$ in the $t_L$-$t_R$ plane, where $\gamma \approx 1$ ($\gamma \approx -1$) denotes $D_L \gg D_R$ ($D_L \ll D_R$), leading to left-localized (right-localized) skin modes. And $\gamma \approx 0$ denotes $D_L \cong D_R$, showing both the left-localized and right-localized skin modes, namely, the bipolar skin effect. Interestingly, the bipolar skin effect cannot emerge under the Abelian condition $[\boldsymbol{d}_L \cdot \boldsymbol{\sigma}, \boldsymbol{d}_R \cdot \boldsymbol{\sigma}] = 0$; consequently, its emergence is only possible under the non-Abelian condition $[\boldsymbol{d}_L \cdot \boldsymbol{\sigma}, \boldsymbol{d}_R \cdot \boldsymbol{\sigma}] \neq 0$, as proved in the Supplementary Material [63]. The proof, which we have further generalized to the two-dimensional case yielding the same conclusion, is also provided therein. Figure 1(f) shows the PBC (solid lines) and OBC (dots) spectra at point $P_3$ within $\gamma \approx 0$. One can see that each PBC energy band winds as two counterclockwise (pink areas) and one clockwise (yellow area) loops, leading to $w = \pm 1$. Consequently, OBC eigenstates exhibit both the left-localized and right-localized skin effects, as shown in Fig. 1(g), which reveals the bipolar skin effect. In addition, the spectral evolution along certain path in Fig. 1(e) is discussed in the Supplementary Material [63].

We implement the non-Abelian Hatano-Nelson model with a nonreciprocal $U(2)$ gauge field in the electric circuit. The photograph of designed circuit with two sites (grey ellipse) is shown in Fig. 2(a). The related circuit schematic is displayed in Fig.

2(b). Each site consists of two circuit nodes, which act as the opposite pseudospin. Here, we provide a universal recipe to achieve arbitrary forms of non-Hermitian matrix-valued couplings. The on-site potential $C_0\sigma_x$ is achieved using a capacitor $C_0$. The matrix-valued couplings are achieved through distinct link configurations of electrical elements, as depicted in Fig. 2(c). Specifically, the leftward coupling $C_1\sigma_z$ is implemented by connecting a capacitor $C_1$ and an inductor $L_1$ in parallel. In the circuit system, the capacitor $C_1$ and inductor $L_1$ host positive admittance coupling $J_{C_1} = i\omega C_1$ and negative admittance coupling $J_{L_1} = -i/\omega L_1$, respectively. Therefore, we choose inductor $L_1$ to achieve negative coupling in $C_1\sigma_z$. While the rightward coupling $C_2\sigma_x$ is realized by capacitor $C_2$ with braided connections. Each node is grounded by a $LCR$ parallel circuit to keep the same on-site terms of each node. Additionally, the resistor $R_0$ is incorporated to enhance the system's stability (Supplementary Material [63]). Based on Kirchhoff's law and the Bloch theorem, we can derive the circuit equation for a site (unit cell): $I_0 = J_0 V_0$, where $I_0$ and $V_0$ represent the input current and response voltage, respectively. $J_0$ is the admittance matrix given by

$$J_0 = i\omega\left[m_0\sigma_0 + m_1(\sigma_0 - \sigma_z)e^{ik} + C_0\sigma_x + C_1 e^{ik}\sigma_z + C_2 e^{-ik}\sigma_x\right], \quad (3)$$

where $m_0 = -C_2 - 2C_0 + 1/\omega^2 L_0 - 1/i\omega R_0 - C_1 + 1/\omega^2 L_1$ and $m_1 = (C_1 - 1/\omega^2 L_1)/2$. Here, the parameter $m_0$ can only cause the movement of admittance spectrum, without altering the topology of the system. When $m_1 = 0$, $\omega = \omega_0 = 1/\sqrt{L_1 C_1}$, the admittance matrix $J_0$ is topologically equivalent to the Hamiltonian $H$ in Eq. (2) with $\boldsymbol{d}_L = (0,0,1)$ and $\boldsymbol{d}_R = (1,0,0)$. Naturally, the admittance spectra $j$ and admittance eigenstates $U$ of $J_0$ take the similar roles to the eigenvalues and eigenstates of $H$. Therefore, we can predict that the admittance spectra will braid as a Hopf link, and the admittance eigenstates will show the bipolar skin effect. The derivation of the admittance matrix and the effect of the extra term $m_1$ are detailed in the Supplementary Material [63].

We first observe the higher complex spectrum braiding with $|\nu| > 1$ in the circuit samples under PBC. Figure 3(a) shows the photograph of the experiment circuit with 19 sites. In the circuit system, the admittance matrix can be measured by the input currents and response voltages [60]. Specifically, under the PBC, an excitation source is placed at each node in a unit cell and measure the corresponding input currents and response voltages of all the nodes (Supplementary Material [63]). By diagonalizing the measured admittance matrix at $\omega_0$, we can observe the admittance spectra in the $E$-$k$ space. Here, we examine three braiding scenarios with braiding degrees $v = -2$, $v = 0$, and $v = 2$, as illustrated in Figs. 3(b)-3(d). Solid lines and dots represent the calculated and measured admittance spectra, respectively, and they exhibit excellent agreement. When $v = 0$, we can observe that the admittance spectral braiding takes the form of an unlink, indicating that the spectra are unbraided. However, when $v = \pm 2$, the admittance spectra manifest as different Hopf links, which goes beyond the situation of traditional 1D non-Hermitian model with nearest-neighbor couplings.

We finally observe the bipolar skin effect in the circuit sample under OBC. In order to demonstrate the existence of the skin effect, we construct a circuit sample with 47 sites and consider two parameter cases. The first case is the monopolar skin effect, where the hopping amplitudes $C_1$ is much smaller than $C_2$. As depicted in Fig. 4(a), the PBC admittance spectra with clockwise loops host negative spectral winding numbers and guarantee the right-localized skin effect. The skin effect can be visualized by the distributions of OBC admittance eigenstates, as shown in Fig. 4(b). The measurement of admittance matrix in the OBC case is petty different from that in the PBC case. Here, the excitation source should be extended to all nodes rather than confined to a unit cell (Supplementary Material [63]). Measured distributions of all OBC admittance eigenstates are plotted in Fig. 4(c), which only exhibit right-localized skin modes and reveal the significant monopolar skin effect. The second case is the bipolar skin effect, where the hopping amplitudes $C_1$ is close to $C_2$. As shown in Fig. 4(d), the PBC spectra with both clockwise and counterclockwise loops possess spectral winding numbers $w = \pm 1$, guaranteeing the presence of the bipolar skin effect. Figures 4(e) and 4(f) show the calculated and measured distributions of all OBC admittance eigenstates, which display a significant bipolar skin effect, with both left- and right-localized skin modes coexisting.

In conclusion, we have theoretically and experimentally reported the non-Hermitian physics, including the spectral braiding with higher braiding degree and the bipolar skin effect, which are aroused by the nonreciprocal $U(2)$ gauge field in a non-Abelian Hatano-Nelson model. Different from the Abelian mechanism of next-nearest-neighbor coupling in a one-band system [13], here the bipolar skin effect is directly attributed to non-Abelian condition. Our work is anticipated to drive forward the research on topological phenomena at the intersection of non-Abelian and non-Hermitian physics. In addition, the designed circuit system can serve as a basis to further explore the exotic non-Hermitian physics and may inspire the development of non-Hermitian circuit devices.

**Acknowledgements**

This work is supported by the National Key R&D Program of China (Grants No. 2022YFA1404900, No. 2022YFA1404500, and No. 2023YFB2804701), National Natural Science Foundation of China (Grants No. 12374409, No. U25D8001, No. 12574484, No. 12574024, and No. 12504250), Natural Science Foundation of Hunan Province (Grant No. 2025JJ60016), and Natural Science Foundation of Education Department of Hunan Province (Grant No. 24B0317).

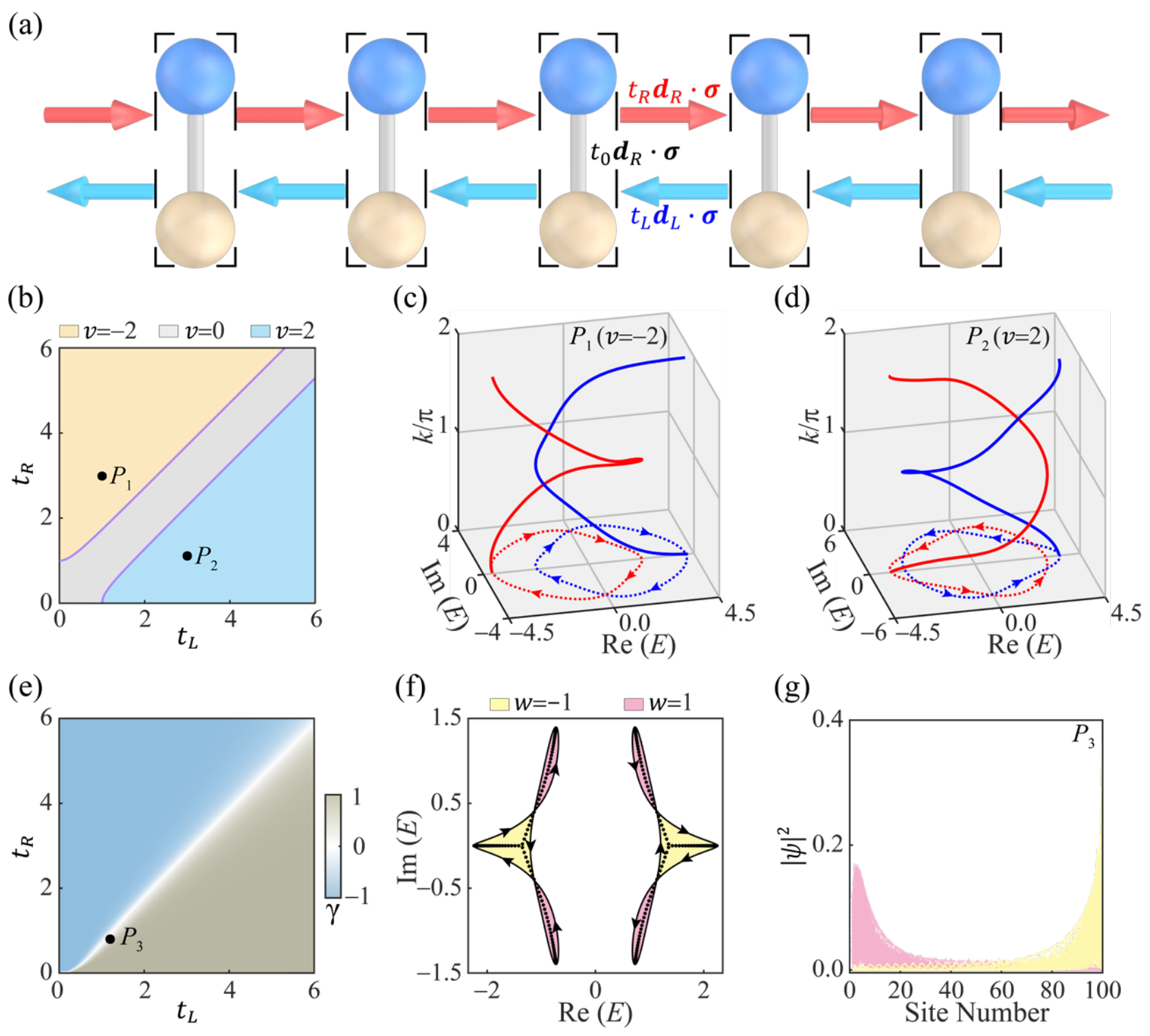

FIG. 1. Hopf-link shaped complex energy braiding and bipolar skin effect in a non-Abelian Hatano-Nelson model with nonreciprocal $U(2)$ gauge field. (a) Schematic of the non-Abelian Hatano-Nelson model, which features on-site potential $t_0\boldsymbol{d_R}\cdot\boldsymbol{\sigma}$ along with nonreciprocal matrix-valued couplings $t_L\boldsymbol{d}_L\cdot\boldsymbol{\sigma}$ and $t_R\boldsymbol{d}_R\cdot\boldsymbol{\sigma}$. The unit cell (black dashed box) contains spin up and down components (blue and light tan spheres). In this work, we choose $\boldsymbol{d}_L=(0,0,1)$ and $\boldsymbol{d}_R=(1,0,0)$. (b) Phase diagram of spectral braiding with $t_0=1$. Phase boundaries (purple lines) denote the appearance of exceptional points. Hopf links occur at phases with braiding degree $v=\pm 2$. (c) and (d) Complex energy braiding in the $E$-$k$ space at the points $P_1$ ($t_L=1,\ t_R=3$) and $P_2$ ($t_L=3,\ t_R=1$) shown in (b). The dashed lines on the $k=0$ plane represent the projections of spectral braiding, and the arrows illustrate the winding direction as $k$ increases. (e) Contrast function $\gamma$ in the $t_L$-$t_R$ plane with $t_0=1$, The bipolar skin effect occurs at the area with $\gamma\approx 0$. (f) PBC (solid lines) and OBC (dots) spectra on the complex plane at the point $P_3$ ($t_L=1.2,\ t_R=0.9$) in (e). The opposite spectral winding numbers $w=1$ (pink areas) and $w=-1$ (yellow areas) indicate the presence of the bipolar skin effect. (g) Distributions of OBC eigenstates at the point $P_3$ in (e), revealing the bipolar skin effect.

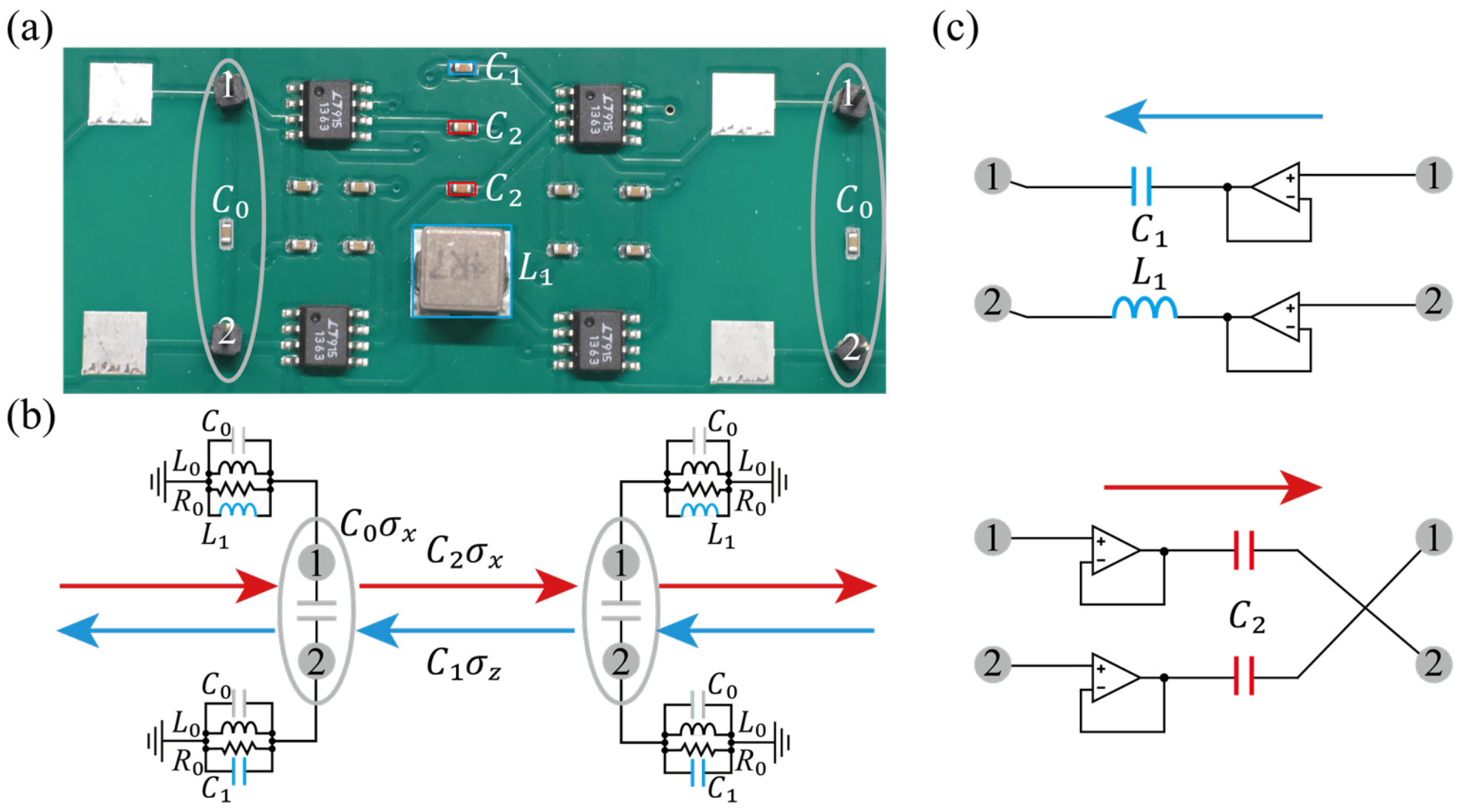

FIG. 2. Circuit implementation of the non-Abelian Hatano-Nelson model with nonreciprocal $U(2)$ gauge field. (a) Photograph of the circuit sample with two sites. Each site contains two circuit nodes 1 and 2. (b) Schematic diagram of the circuit design. The onsite potential is achieved by capacitor $C_0$. (c) Link configurations of circuit design, which implement nonreciprocal $U(2)$ couplings $C_1\sigma_z$ (top) and $C_2\sigma_x$ (bottom).

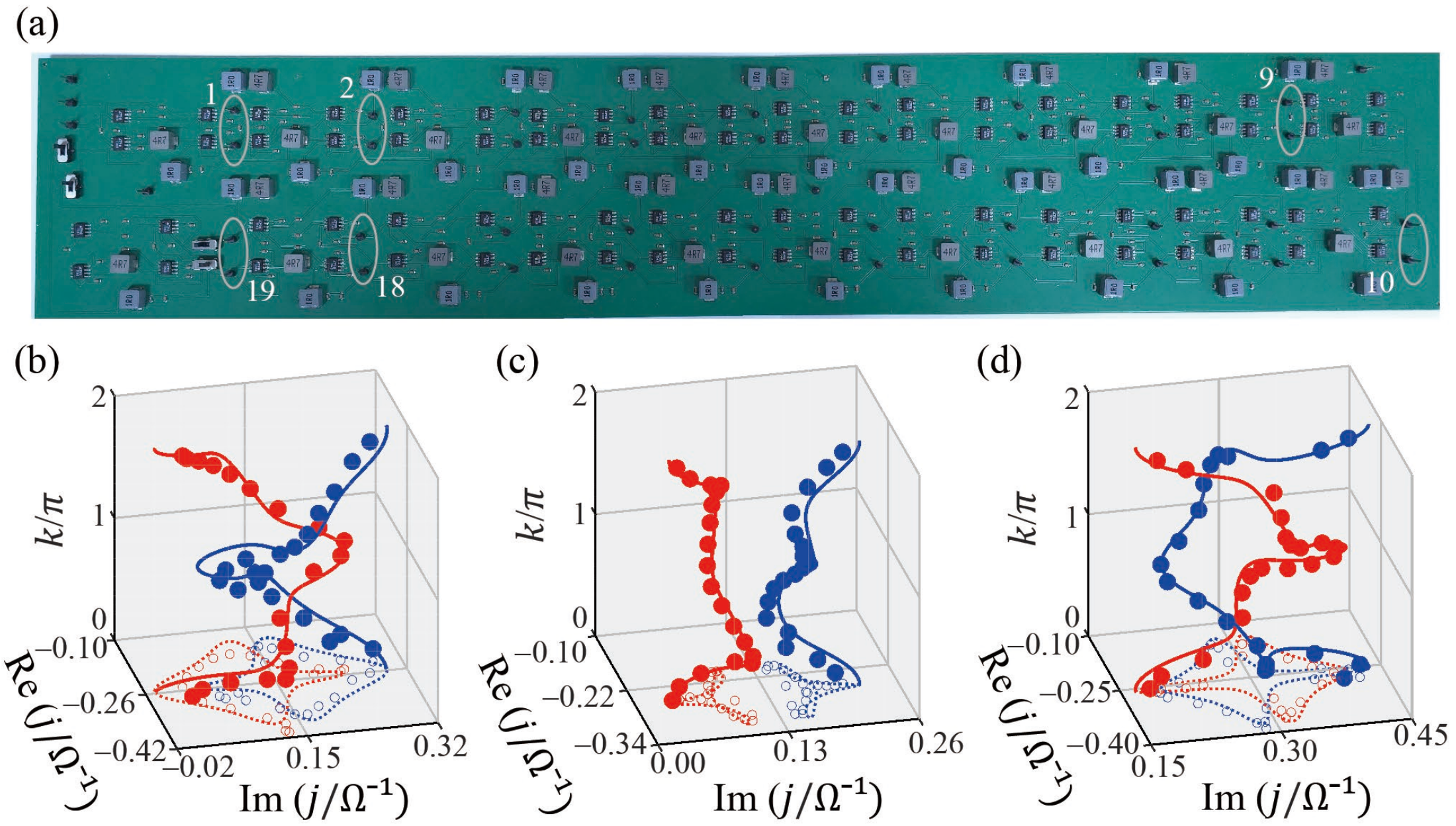

FIG. 3. Observation of the complex spectrum braiding in the electric circuit. (a) Photograph of the circuit sample with 19 sites under PBC. (b)-(d) Calculated (solid lines) and measured (solid dots) admittance spectra ($j$) for the braiding degrees $v = -2$, $v = 0$, and $v = 2$. Dashed lines and hollow dots on the $k = 0$ plane represent the projection of calculated and measured results, respectively. The parameters are chosen as $C_1 = 20$ nF and $C_2 = 30$ nF in (b), $C_1 = 12$ nF and $C_2 = 9$ nF in (c), $C_1 = 39$ nF and $C_2 = 30$ nF in (d). The other parameters are chosen as $C_0 = 10$ nF, $L_0 = 0.95$ μH, $L_1 = 4.4$ μH and $R_0 = 3.9$ Ω.

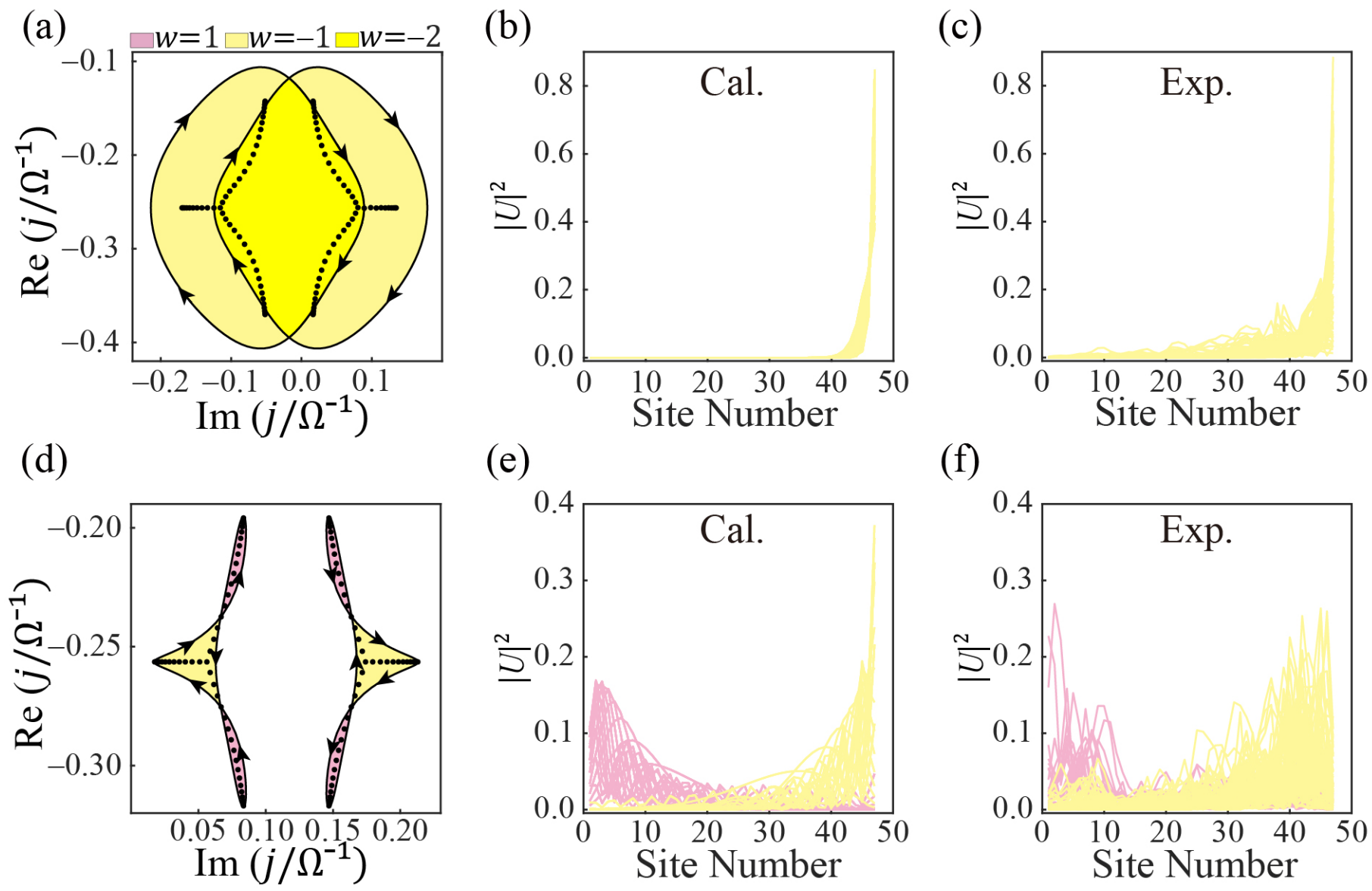

FIG. 4. Observation of the skin effect in the electric circuit. (a) Calculated PBC (solid lines) and OBC (dots) admittance spectra ($j$) on the complex plane. The spectral winding number $w < 0$ indicates the presence of only the monopolar skin effect localized at the right boundary. (b) and (c) Calculated and measured distributions of OBC admittance eigenstates ($U$) in (a). (d) Calculated PBC and OBC admittance spectra on the complex plane. The coexistent of opposite spectral winding numbers $w = 1$ and $w = -1$ indicates the presence of the bipolar skin effect. (e) and (f) Calculated and measured distributions of OBC admittance eigenstates in (d). The parameters are chosen as $C_1 = 10$ nF and $C_2 = 30$ nF in (a)-(c), $C_1 = 12$ nF and $C_2 = 9$ nF in (d)-(f). The other parameters are chosen as $C_0 = 10$ nF, $L_0 = 0.95$ μH, $L_1 = 4.4$ μH and $R_0 = 3.9$ Ω.